# On The Optimization of Dijkstra's Algorithm

Seifedine Kadry, Ayman Abdallah, Chibli Joumaa

American University oft he Middle East, Egaila, Kuwait
{seifedine.kadry, Ayman.abdallah, Chibli.joumaa}@acm.edu.kw

**Abstract.** In this paper, we propose some amendment on Dijkstra's algorithm in order to optimize it by reducing the number of iterations. The main idea is to solve the problem where more than one node satisfies the condition of the second step in the traditional Dijkstra's algorithm. After application of the proposed modifications, the maximum number of iterations of Dijkstra's algorithm is less than the number of the graph's nodes.

## 1 Introduction

The shortest-route problem determines a route of minimum weight connecting two specified vertices, source and destination, in a weight graph (digraph) in a transportation network. Other situations can be represented by the same model like VLSI design, equipment replacement and others. There are different types of shortest path algorithm [5] to find the shortest path of any graph. Most frequently encountered are the following:
- Shortest path between two specified vertices
- Shortest path between all pairs of vertices
- Shortest path from a specified vertex to all others

The most efficient algorithm used to find the shortest path between two known vertices is Dijkstra's algorithm [1]. Some improvements on Dijkstra algorithm are done in terms of efficient implementation [3] and cost matrix [4]. In this paper, we propose some improvement in order to reduce the number of iterations and to find easily and quickly the shortest path.

## 2 Dijkstra's Algorithm

The problem of finding the shortest path from a specified vertex s to mother t can be stated as follows:
A simple weighted digraph G of n vertices is described by a n by n matrix $D=[d_{ij}]$, where, $d_{ij}$ = length (or distance or weight) of the directed edge from vertex i to vertex j:

$$d_{ij} = \begin{cases} > 0, & \text{if } i \neq j \\ = 0, & \text{if } i = j \\ = \infty \ (large\ number), & \text{if there is no edge from } i \text{ to } j \end{cases}$$

Dijkstra's algorithm labels the vertices *of* the given digraph, At each stage in the algorithm some vertices have permanent labels and others temporary labels. The algorithm begins by assigning a permanent label 0 to the starting vertex *s,* and temporary label infinity to the remaining n-1 vertices. From then on in each iteration, another vertex sets a permanent label, according to the following rules:

    **a.** Every vertex j that is not yet permanently labeled gets a new temporary label whose value is given by min[old label of j, (old label of i + $d_{ij}$)], where i is the latest vertex permanently labeled, in the previous iteration, and $d_{ij}$ is the direct distance between vertices i and j. If i and j are not joined by an edge, then $d_{ij}$=infinity.

    **b.** The smallest value among all the temporary labels is found, and this becomes the permanent label of the corresponding vertex. In case of more than one shortest path, select any one of the candidates for permanent labeling. Steps a and b are repeated alternately until the destination vertex t gets a permanent label. The first vertex to be permanently labeled is at a distance of zero from s. The second vertex to get a permanent label (out of the remaining n-1 vertices) is the vertex closest to s From the remaining n-2 vertices, the next one to be permanently labeled is the second closest vertex to s. And so on. The permanent label of each vertex is the shortest distance of that vertex from s. This statement *can* be proved by induction.

## 3  Modified Dijkstra's Algorithm

In this section, we will present a modified version of Dijkstra's Algorithm in order to reduce the total number of iteration by optimized the situation of many shortest paths:

    a. Every vertex j that is not yet permanently labeled gets a new temporary label whose value is given by min[old label of j, (old label of i + $d_{ij}$)], where i is the latest vertex permanently labeled, in the previous iteration, and $d_{ij}$ is the direct distance between vertices i and j. If i and j are not joined by an edge, then $d_{ij}$=infinity [6].

    b. The smallest value among all the temporary labels is found, and this becomes the permanent label of the corresponding vertex. In case of more than one shortest path, select all of them for permanent labeling.

    c. Steps (a) and (b) are repeated alternately maximum n-1 times until the destination vertex t gets a permanent label.

    d. Le T=[$t_{ij}$] is the shortest path matrix that we can build it form D, as follows

$$t_{ij} = \begin{cases} 0, & \text{if } d_{ji} = 0 \text{ or } d_{ji} = \infty \\ d_{ji}, & \text{otherwise} \end{cases}$$

## 3  Illustrative Example

The network in the following figure (figure 1) gives the distances in miles between pairs of cities 1,2,… and 8. For efficiency purpose, we will find the shortest route between cities 1 and 8 using the traditional and modified version of Dijkstra algorithm:

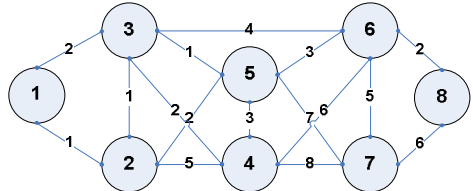

Fig. 1: Network of 8 nodes

*Dijkstra algorithm:*

The number of iterations to find the shortest route between cities 1 and 8 by using Dijkstra algorithm is 8 (figure 2), this is the result of TORA [1] software:

| Node | Label | Status |
|---|---|---|
| **Iteration 1** | | |
| 1 | [0.00, --] | permanent |
| 2 | [1.00, 1] | temporary |
| 3 | [3.00, 1] | temporary |
| 4 | | |
| 5 | | |
| 6 | | |
| 7 | | |
| 8 | | |
| **Iteration 2** | | |
| 1 | [0.00, --] | permanent |
| 2 | [1.00, 1] | permanent |
| 3 | [2.00, 2] | temporary |
| 4 | [6.00, 2] | temporary |
| 5 | [3.00, 2] | temporary |
| 6 | | |
| 7 | | |
| 8 | | |
| **Iteration 3** | | |
| 1 | [0.00, --] | permanent |
| 2 | [1.00, 1] | permanent |
| 3 | [2.00, 2] | permanent |
| 4 | [4.00, 3] | temporary |
| 5 | [3.00, 2] | temporary |
| 6 | [6.00, 3] | temporary |
| 7 | | |
| 8 | | |
| **Iteration 4** | | |
| 1 | [0.00, --] | permanent |
| 2 | [1.00, 1] | permanent |
| 3 | [2.00, 2] | permanent |
| 4 | [4.00, 3] | temporary |
| 5 | [3.00, 2] | permanent |
| 6 | [6.00, 3] | temporary |
| 7 | [10.00, 5] | temporary |
| 8 | | |
| **Iteration 5** | | |
| 1 | [0.00, --] | permanent |
| 2 | [1.00, 1] | permanent |
| 3 | [2.00, 2] | permanent |
| 4 | [4.00, 3] | permanent |
| 5 | [3.00, 2] | permanent |
| 6 | [6.00, 3] | temporary |
| 7 | [10.00, 5] | temporary |
| 8 | | |
| **Iteration 6** | | |
| 1 | [0.00, --] | permanent |
| 2 | [1.00, 1] | permanent |
| 3 | [2.00, 2] | permanent |
| 4 | [4.00, 3] | permanent |
| 5 | [3.00, 2] | permanent |
| 6 | [6.00, 3] | permanent |
| 7 | [10.00, 5] | temporary |
| 8 | [8.00, 6] | temporary |
| **Iteration 7** | | |
| 1 | [0.00, --] | permanent |
| 2 | [1.00, 1] | permanent |
| 3 | [2.00, 2] | permanent |
| 4 | [4.00, 3] | permanent |
| 5 | [3.00, 2] | permanent |
| 6 | [6.00, 3] | permanent |
| 7 | [10.00, 5] | temporary |
| 8 | [8.00, 6] | permanent |
| **Iteration 8** | | |
| 1 | [0.00, --] | permanent |
| 2 | [1.00, 1] | permanent |
| 3 | [2.00, 2] | permanent |
| 4 | [4.00, 3] | permanent |
| 5 | [3.00, 2] | permanent |
| 6 | [6.00, 3] | permanent |
| 7 | [10.00, 5] | permanent |
| 8 | [8.00, 6] | permanent |

Fig. 2: Iterations 1 to 8 of Dijkstra algorithm

*Modified Dijkstra algorithm:*

By using the modified version of Dijkstra algorithm, 5 iterations only are required to get the shortest route between cities 1 and 8:

Initially

$$D = \begin{bmatrix} 0 & 1 & 2 & \infty & \infty & \infty & \infty & \infty \\ 1 & 0 & 1 & 5 & 2 & \infty & \infty & \infty \\ 2 & 1 & 0 & 2 & 1 & 4 & \infty & \infty \\ \infty & 5 & 2 & 0 & 3 & 6 & 8 & \infty \\ \infty & 2 & 1 & 3 & 0 & 3 & 7 & \infty \\ \infty & \infty & 4 & 6 & 3 & 0 & 5 & 2 \\ \infty & \infty & \infty & 8 & 7 & 5 & 0 & 6 \\ \infty & \infty & \infty & \infty & \infty & 2 & 6 & 0 \end{bmatrix}$$

and the set of permanent vertices is P={1}.

| Iteration 1: | $P = \{1,2\}, d_{1j} = \{0,1,2,4,\infty,\infty,\infty,\infty\}$ |
|---|---|
| Iteration 2: | $P = \{1,2,3\}, d_{1j} = \{0,1,2,4,3,6,\infty,\infty\}$ |
| Iteration 3: | $P = \{1,2,3,5\}, d_{1j} = \{0,1,2,4,3,6,10,\infty\}$ |
| Iteration 4: | $P = \{1,2,3,4,5,6\}, d_{1j} = \{0,1,2,4,3,6,10,8\}$ |
| Iteration 5: | $P = \{1,2,3,4,5,6,7,8\}, d_{1j} = \{0,1,2,4,3,6,10,8\}$ |

$$T = \begin{bmatrix} 0 & 1 & 0 & 0 & 0 & 0 & 0 & 0 \\ 0 & 0 & 1 & 0 & 2 & 0 & 0 & 0 \\ 0 & 0 & 0 & 2 & 0 & 4 & 0 & 0 \\ 0 & 0 & 0 & 0 & 0 & 0 & 0 & 0 \\ 0 & 0 & 0 & 0 & 0 & 0 & 7 & 0 \\ 0 & 0 & 0 & 0 & 0 & 0 & 0 & 2 \\ 0 & 0 & 0 & 0 & 0 & 0 & 0 & 0 \\ 0 & 0 & 0 & 0 & 0 & 0 & 0 & 0 \end{bmatrix}$$

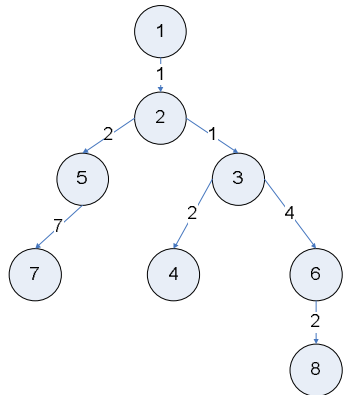

Fig. 3: Tree of shortest route

Based on the previous tree (figure 3), the shortest route and distance between cities 1 and 8 are: 1-2-3-6-8 and the distance is 8 miles.

## 4 Conclusion

In this paper an optimization of Dijkstra algorithm is presented. The main idea is to deal with the second stage of the traditional version of this algorithm where the shortest routes from a node to its successors are many and to reduce the number of iterations. The main feature of the proposed amendment is the remarkable reduction in the number of iterations and the ease in finding the shortest route from any city to any city. In the future work, we will implement the proposed modifications then to compare it with the existing one in terms of complexity.